# When can we get away with using the two-way fixed effects regression?

**Apoorva Lal**

*Netflix, Los Gatos, CA*

## Abstract

The use of the two-way fixed effects regression in empirical social science was historically motivated by folk wisdom that it uncovers the average treatment effect on the treated (ATT). This has come under scrutiny recently due to recent results in applied econometrics showing that it fails to uncover meaningful averages of heterogeneous treatment effects in the presence of effect heterogeneity over time and across adoption cohorts, and several heterogeneity-robust alternatives have been proposed. However, these estimators often have higher variance and are therefore under-powered for many applications, which poses a bias-variance tradeoff that is challenging for researchers to navigate. In this paper, we propose simple tests of linear restrictions that can be used to test for differences in dynamic treatment effects over cohorts, which allows us to test for when the two-way fixed effects regression is likely to yield biased estimates of the ATT. These tests are implemented as methods in the pyfixest python library.

## 1 Introduction

Difference-in-Differences and event-study are now the most popular for estimating causal effects in observational settings thanks to the growing importance of administrative and other offline data sources ([1]). Their popularity arises from their broad applicability to settings with selection on unobservable unit-specific factors, and straightforward implementation as a two-way fixed effects linear regression in the two-period case. As with many empirical techniques, however, practice outstrips theory, and the extension of the equivalence between fixed-effects regression and the non-parametric estimator in the two-period, two-cohort case turned out to be more complicated than previously realized, which has prompted an exposion of alternative estimators in that are robust to the 'contamination bias' introduced into the two-way fixed-effects regression by treatment effect heterogeneity. Since these newer estimators either trim the data or add many parameters to zero out the bias, they tend to have considerably higher variance, which introduces a bias-variance tradeoff that is challenging to navigate for practitioners. In this paper, we propose the application of classical joint-tests of appropriately parametrized linear regressions as a tool to aid practitioners in navigating the bias-variance tradeoff in difference-in-differences settings.

## 2 Methodology

Consider a balanced panel-data setting with $i = 1, ..., N$ individuals observed over $t = 1, ..., T$ time periods. For each unit $i$, a binary treatment $w_{it} := 1(t \geq g_i)$ is assigned at some adoption time $g_i \in \mathcal{G}$ where $\mathcal{G} := [T] \cup \infty$ is the set of treatment adoption times and $g_i = \infty$ indicates a never-treated unit. We observe a scalar outcome $y_{it} = w_{it} y_{it}^1 + (1 - w_{it}) y_{it}^0$, where $y_{it}^1$ and $y_{it}^0$ are potential outcomes under treatment and control, respectively[1]. The following two-way fixed effects regression

$$y_{it} = \tau w_{it} + \alpha_i + \lambda_t + \varepsilon_{it} \tag{1}$$

is a workhorse regression in applied economics and adjacent fields for the estimation of causal effects in such settings. The estimand that researchers typically seek to estimate in panel data settings is the Average Treatment effect on the Treated (ATT) ($\mathbb{E}[y_{it}^1 - y_{it}^0 \mid w_{it} = 1]$), and researchers often interpret the coefficient on the treatment indicator, $\hat{\tau}$, as an estimate of the ATT. The above regression's dynamic ('event study') counterpart

$$y_{it} = \sum_{s \neq -1}^{T} \gamma_s \Delta_{it}^s + \alpha_i + \lambda_t + \varepsilon_{it} \tag{2}$$

where $\Delta_{it}^s$ is an indicator for the $s$-th period relative to the adoption time for treated units (which in turn is the first-difference of the treatment indicator, [2]). The presence of leads and lags of the switching indicator in this regression allows us to interpret the coefficients on lags as estimate the dynamic ATT ([3] ch 5) and coefficient on leads as a visual check of the validity of the parallel trends. This practice is widespread in applied research but tends to be distortionary and has low power ([4]).

---

[1]Defining potential outcomes as $y_{it}^w$ is a strong but common assumption; it requires no carryover - that the outcome for unit $i$ at time $t$ is only influenced by $i$'s current-period treatment and not treatment history. Alternative estimators

When $g_i \in \{T_0, \infty\}$ (one-shot adoption), the above regressions are unbiased estimates of the ATT under the assumption of parallel trends and no anticipation ([5]). However, when $g_i \in \{T_0, ..., T-1\}$ (staggered adoption), the above regressions exhibit the 'negative weighting'/'contamination bias' problem ([6], [7], [8]). As in the cross-sectional case, the regression coefficient on the treatment indicator, $\hat{\tau}$, is a weighted average of the treatment effects over time and across treated cohorts, where the weights are functions of the conditional variance in the treatment. Unlike in the cross-sectional case, however, these weights can be negative for some cohorts, which yields the conclusion that the two-way fixed effects regression can fail to uncover meaningful averages of heterogeneous treatment effects over time and across adoption cohorts[2]. The same is true for the event study coefficient vector $\gamma$ ([10]).

This has prompted a explosion of research in applied econometrics on new estimators that aim to uncover the ATT in the presence of heterogeneous treatment effects over time and across adoption cohorts ([9], [11], [12] for reviews). Such heterogeneity-robust estimators typically involve estimating the ATT separately for each cohort using tailored comparisons between each treated cohort and either a never-treated or not-yet-treated group, and then averaging (optionally weighted by inverse-propensity weights, e.g. [13]) these estimates to obtain an overall estimate of the ATT. While their consistency properties for the ATT are well understood and they avoid the negative weighting problem by construction, they are often computationally expensive and have higher variance than the two-way fixed effects regression.

This poses a practical bias-variance tradeoff for researchers: while the two-way fixed effects regression is computationally simple and has low variance, it may yield biased estimates of the ATT in the presence of heterogeneous treatment effects over time and across adoption cohorts. In contrast, heterogeneity-robust estimators are computationally expensive and have higher variance,

---

such as Marginal Structural Models (MSMs) and dynamic panel models permit estimation in the presence of carryover under different strong assumptions but are considerably more computationally challenging, and as such are used infrequently.

[2]In particlar, this constitutes a violation of the 'no-sign reversal property' where $\hat{\tau}$ is positive even if the treatment effect is strictly negative for each $(g, t)$ ([9]).

but they are consistent for the ATT in the presence of heterogeneous treatment effects over time and across adoption cohorts. As a practical matter, a large re-analysis of published work in political science by [14] finds that they rarely overturn the conclusions of the two-way fixed effects regression, and are typically have considerably larger variance. Similarly, [15] finds that most new heterogeneity-robust estimators are underpowered for realistic effect sizes in the state-level US setting where difference-in-differences approaches commonly used.

This motivates the primary focus of this paper: to develop simple tests that can be used to test for differences in dynamic treatment effects over cohorts, which allows us to test for when the two-way fixed effects regression is likely to yield biased estimates of the ATT. Heuristically, if the dynamic treatment effects are homogeneous over cohorts, then the two-way fixed effects regression is likely to yield unbiased estimates of the ATT that are considerably more precise than alternative estimators that typically discard more data in order to shut down the negative weighting problem.

To build intuition for this approach, consider Figure 1 and Figure 2. In Figure 1, there are three adoption cohorts (plus a never-treated cohort - bottom panel), and all cohorts exhibit the same temporal heterogeneity pattern (the effect function is $\log(t)$ - top panel), and so the 2WFE event study (blue line in panel 2) is consistent for the true dynamic ATT (black line in panel 2). We can also consistently estimate the cohort-level ATTs with an appropriately saturated regression ([10], [16]) as shown in the third panel. In Figure 2, in contrast, we have the same three adoption cohorts, but the three cohorts exhibit radically different temporal heterogeneity: the first exhibits a linear decay down to zero, the second exhibits a log increase followed by zero, and the third exhibits sinusoidal effects. In this case, the 2WFE event study (blue line in panel 2) is not consistent for the true dynamic ATT (black line in panel 2); in fact, the estimated event study suggests a violation of the parallel trend assumption despite the treatments being randomized and thus parallel trends being true in the DGP, which is a pernicious side-effect of the negative weights problem. We can still estimate the cohort-level ATTs correctly with a saturated regression. The key insight is that testing for differences between a 'pooled' event study (the blue line in the second panel) and cohort X time interactions (that yield the cohort-level estimates in the third panel) can help us distinguish between the two scenarios. This can be formulated as a joint F-test on the coefficients of the cohort X time interactions in a saturated regression. We provide a formal statement of this test in the next section, and show through simulation studies that this approach can detect cohort-level temporal heterogeneity in a variety of DGPs.

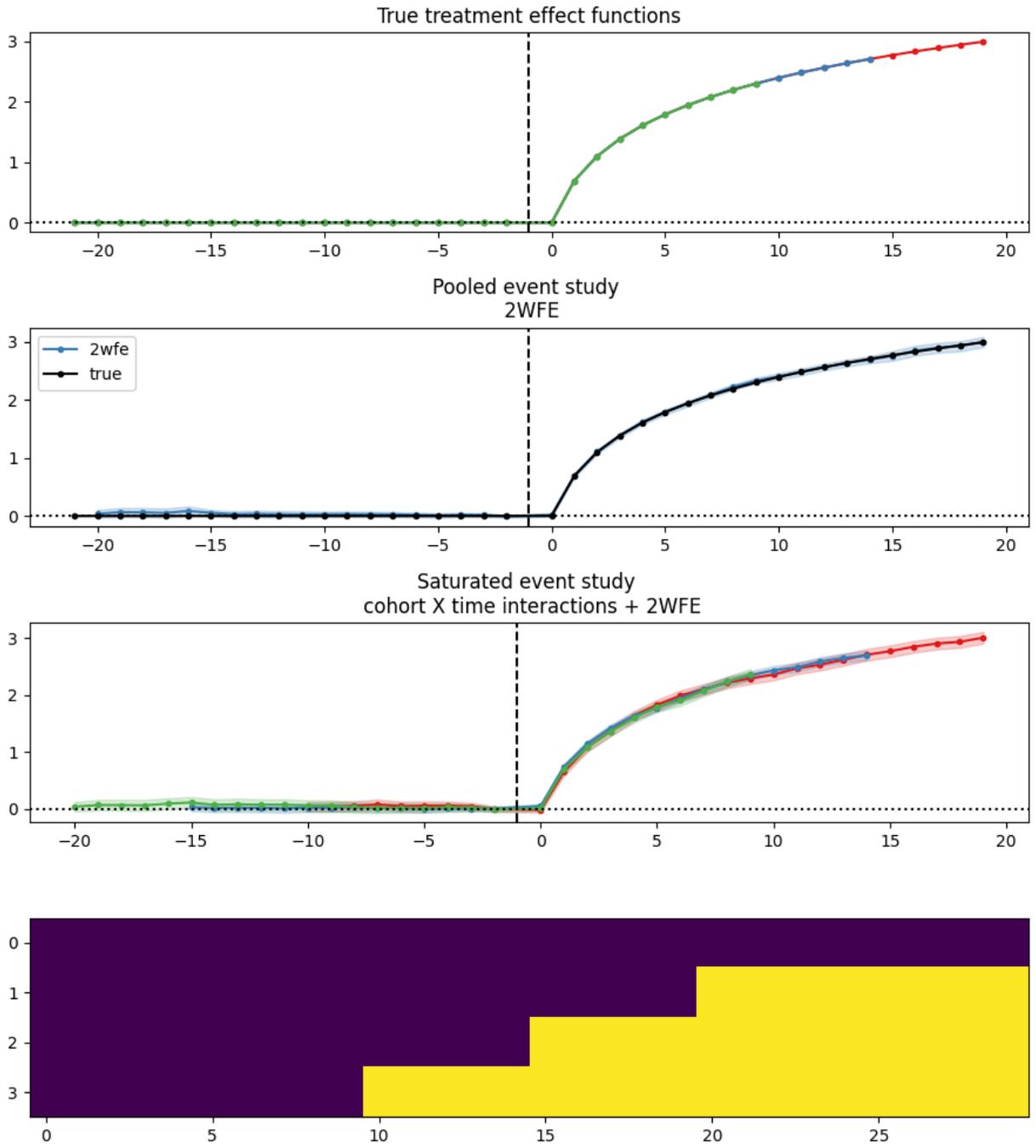

Figure 1: true and estimated effects from pooled and saturated event study regressions with homogeneous treatment effects across three cohorts. Joint test p-value = 0.11

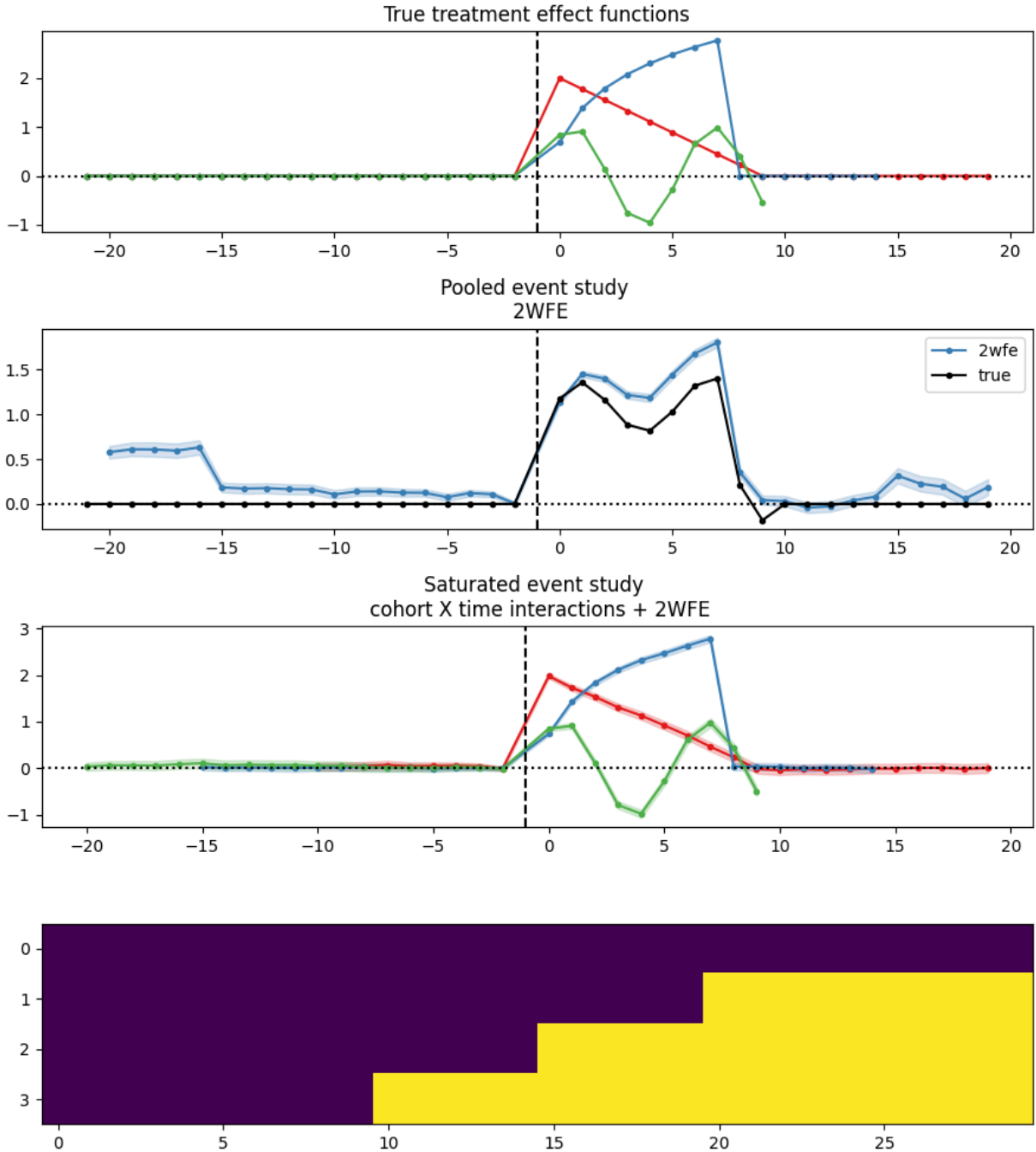

Figure 2: true and estimated effects from pooled and saturated event study regressions in a DGP with heterogeneous treatment effects across three cohorts. Joint test p-value = 0.000

## 3 Methodology

Tests considered in the following section take the form of traditional joint tests of multiple linear restrictions, where the null hypothesis is that $R\beta = q$ where $R$ is a $m \times k$ matrix of linear restrictions, $\beta$ is a $k \times 1$ vector of coefficients, and $q$ is a $m \times 1$ vector of constants. The test statistic is then

$$F = \frac{\left(\boldsymbol{R}\hat{\beta} - \boldsymbol{q}\right)'\left[\boldsymbol{R}\hat{\mathbb{V}}\boldsymbol{R}'\right]^{-1}\left(\boldsymbol{R}\hat{\beta} - \boldsymbol{q}\right)}{m} \sim F(m, n-k) \text{ under the null hypothesis} \quad (3)$$

where $\hat{\mathbb{V}}$ is the cluster-robust variance-covariance matrix of the coefficient estimates. We consider two tests: one for testing for event study dynamics, and one for testing for heterogeneity in event study dynamics. These tests are both classical Wald tests for linear restrictions and are asymptotically equivalent to the Likelihood Ratio test and Lagrange Multiplier test. The test is optimal (most powerful) in the class of invariant tests for local alternatives when errors are normally distributed ([17] lemma 8.5.2).[3]

### 3.1 Testing for event study dynamics

As a warmup, consider a simple comparison between (1) and (2). The latter decomposes the ATT across time-periods. For the purposes of testing for event study dynamics, we only care about comparing the equality of the dynamic treatment effects after the treatment is assigned ($\{\gamma_t\}_{t=0}^T$) against the common ATT estimate $\tau$. We can test the following null hypothesis

$$H_0 : \{\gamma_t\}_{t=0}^T = \hat{\tau} \text{ for all k} > 0 \quad (4)$$

by specifying $\boldsymbol{R} = \boldsymbol{I}_K$ as a $T_1 \times T_1$ identity matrix and $\boldsymbol{q} = (\hat{\tau}, ..., \hat{\tau})'$ as a $T_1$-vector of the restricted estimate ($\hat{\tau}$ from (1)).[4]

### 3.2 Testing for across-cohort heterogeneity in dynamic treatment effects

Next, we extend the approach outlined above to construct a test for across-cohort heterogeneity in dynamic treatment effects. A conventional method to estimate the cohort-level ATTs is to estimate the dynamic treatment effects separately for each cohort and then average these estimates to obtain an overall estimate of the ATT ([10], [16], [18]), which involves specifying the following regression

$$y_{it} = \alpha_i + \lambda_t + \underbrace{\sum_{g_i \in \mathcal{C} \setminus \infty} \sum_{s \neq -1}^T \mathbb{1}(g_i = c) \tau^{sc} \Delta_{it}^s}_{\text{Cohort-Time Interactions}} + \varepsilon_{it} \quad (5)$$

This is a saturated event study that constructs cohort × time interactions for each adoption cohort (with $g_i = \infty$ never treated cohort) omitted and therefore recovers the cohort-level event studies. These coefficients are reported in the third panel in Figure 1 and Figure 2, and correctly uncover the true cohort-level ATTs in the presence of arbitrary heterogeneous treatment effects across cohorts (top panel). The downside of this approach, however, are twofold. First, these regressions can get unwieldy with many cohorts, and the number of parameters grows linearly with the number of cohorts. Second, the cohort level ATTs are self-contained and therefore constructing a test for equality across multiple cohorts is not straightforward. Instead, one may re-specify the saturated event-study regression (5) as follows:

$$y_{it} = \alpha_i + \lambda_t + \underbrace{\sum_{s \neq -1}^T \gamma_s \Delta_{it}^s}_{\text{(a) Common event study coefficients}} + \underbrace{\sum_{c \in \mathcal{C}} \sum_{s \neq -1}^T \delta_s \Delta_{it}^{cs}}_{\text{(b) Cohort-specific deviations}} + \varepsilon_{it} \quad (6)$$

---

[3]This can be implemented using either a $\chi^2$ or $F$ test; the distinction between the two is due to different degrees of freedom that disappear for realistic sample sizes

[4]this can equivalently be formulated by testing for the equality of adjacent elements of $\gamma$, e.g. $\gamma_1 = \gamma_2$ by specifying $\boldsymbol{R}$ that contains rows like $[1, -1, 0, ..., 0]$ and $q = [0, ..., 0]$.

(6) returns numerically identical estimates of the cohort-level dynamic ATT as (5), but it allows us to test for differences in dynamic treatment effects over cohorts more easily. This is because (6) contains a common event study coefficient vector (a), and cohort-level deviations (b). The (b) terms can be jointly tested against the null of zero, which serves as a direct test of cohort-level treatment effect heterogeneity relative to a traditional event study. This approach is similar to omnibus tests of effect heterogeneity in cross-sectional RCTs proposed by [19], testing the joint null of $\gamma = 0$ in the interacted regression $y \sim \tau W + X\beta + WX\gamma + \varepsilon$ serves as a test for explained effect heterogeneity. We illustrate an application of this test in Figure 3, where the top panel reports the saturated event study (5), the middle panel reports the coefficients from re-specified model (6), and the bottom panel reports the sum of the common event study and cohort-specific deviations, which reproduces the saturated event study estimates exactly.

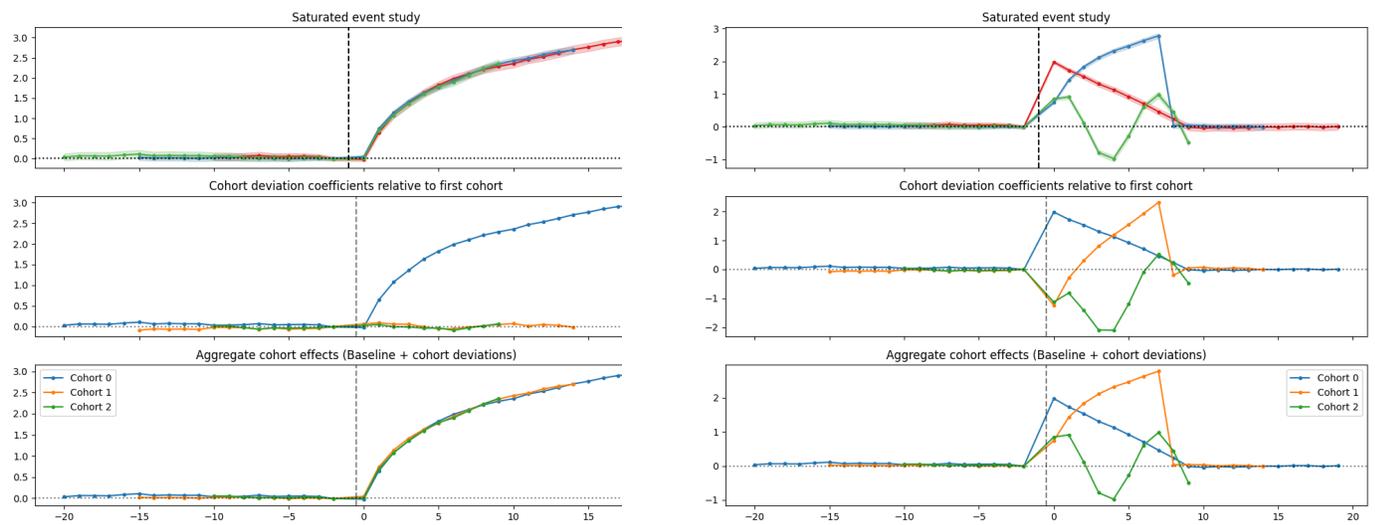

Figure 3: For each DGP (homogeneous - Figure 1 - on the left and heterogeneous - Figure 2 - on the right), the top panel illustrates the traditional event study estimates from eqn (5), which are unbiased for the true effects. The middle panel plots the re-specified model, which plots an overall event study (first cohort : blue) and subsequent cohort deviations (second and third cohorts - which are null in this DGP). The final panel plots the sum of the blue and cohort-specific coefficients, which reproduces the event study coefficient from the first panel exactly.

We show in the next section that this test is consistent for the null hypothesis of homogeneous dynamic treatment effects over cohorts, and that it has power against a variety of alternatives. As a concrete example, the joint $p-$value for the cohort $\times$ time interactions in Figure 1 is $0.11$, while the joint p-value for the cohort $\times$ time interactions in Figure 2 is $0.000$. Thus, we can reject the null hypothesis of homogeneous dynamic treatment effects in Figure 2 but not in Figure 1, which is consistent with the underlying DGP. In the next section, we show through simulation studies that this test has good power to detect across-cohort heterogeneity in dynamic treatment effects in a variety of DGPs.

## 4 Simulation Studies

### 4.1 Testing for event study dynamics

To begin, we perform simulation studies based on to study the properties of the testing procedure described in Section 3.1. We consider the simple setting with a single adoption cohort where the treatment effects follow one of the following seven DGPs visualised in Figure 4.

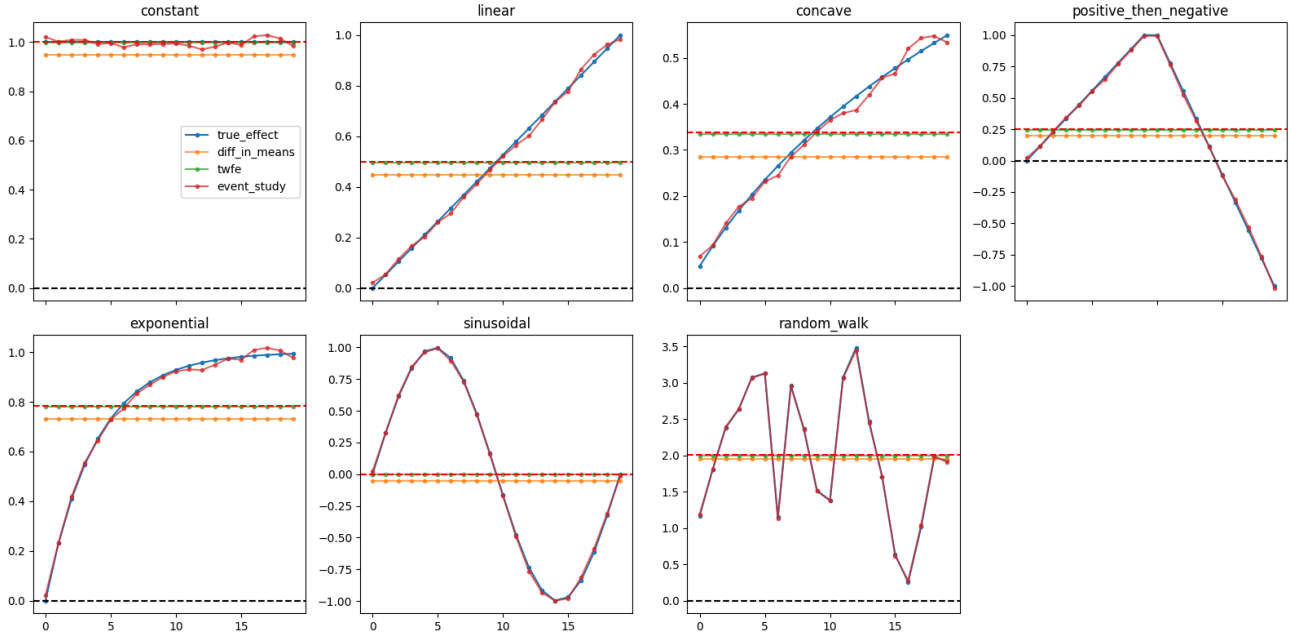

Figure 4:  true treatment effect functions and estimates from difference in means, static, and dynamic two-way fixed effects regressions. The treatment effect is truly stationary in the first DGP and varies over time in the others.

The first DGP has constant effects over time, while the others have varying degrees of temporal heterogeneity. We simulate 1000 replications of the data for each DGP, and compute the rejection rate of the joint test for dynamic treatment effects outlined in the previous section. We report the rejection rate and p-value distribution in Figure 5. We find that the rejection rate for the constant DGP (null) is under the nominal level of $\alpha = 0.05$, while the rejection rates for the other DGPs considerably higher. The rejection rate for concave effects is considerably lower, although this is likely due to the fact that the treatment effects do actually tail off in later time periods and the static effect captures this well.

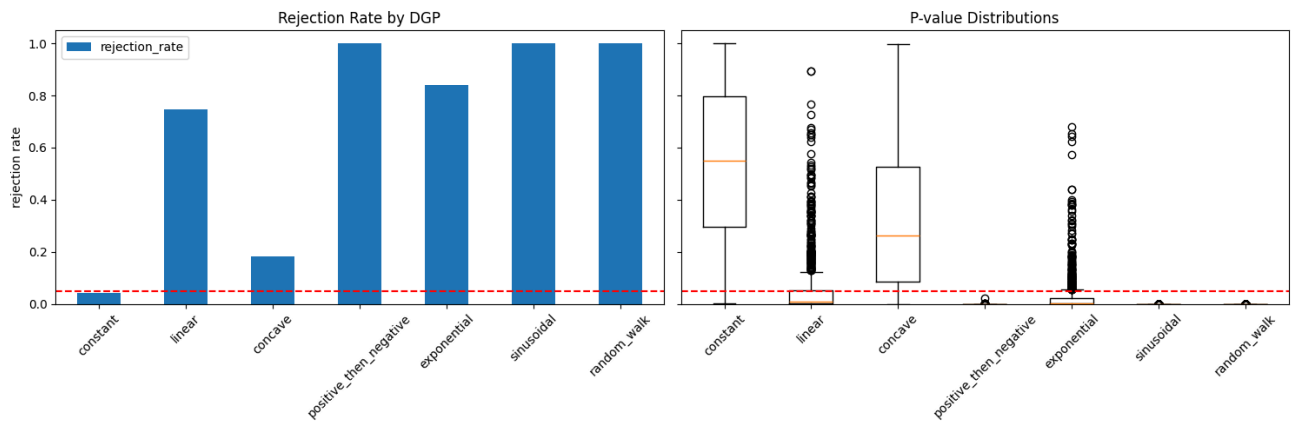

Figure 5: Rejection rates over 1000 replications for the joint test of dynamic treatment effects using an F-test in DGPs from Figure 4

## 4.2 Testing for across-cohort heterogeneity in dynamic treatment effects

Next, we perform simulation studies based on to study the properties of the testing procedure described in Section 3.2. Here, we consider seven different DGPs with homogeneous and heterogeneous treatment effect functions across cohorts as illustrated in Figure 6. In addition to the two DGPs described in the previous section, we consider DGPs with heterogeneity that applies a scaler multiplier to the concave (log) effect function in Figure 1 with 'small' and 'large' differences; a DGP with 'selection on gains' where the cohort with the largest treatment effect adopts first; a DGP with 'novelty effects' where the treatment effect is large for the first few periods and then diminishes; and finally a DGP with 'activity bias' where the treatment effect is immediate and large for the earliest adopting cohort and much more gradual for the others. Among all these DGPs, the homogenous and novelty effects DGPs have homogeneous treatment effects across cohorts, while all others have heterogeneous treatment effects across cohorts.

For each DGP, we simulate 1000 replications of the data, and compute the rejection rate of the joint test for cohort-level coefficients outlined in the previous section. We report the rejection rate and p-value distribution in Figure 7. We find that the rejection rate for the homogeneous DGP (null) is under the nominal level of $\alpha = 0.05$, while the rejection rates for heterogeneous DGPs are close to 1. This suggests that the test has good power to detect across-cohort heterogeneity in dynamic treatment effects.

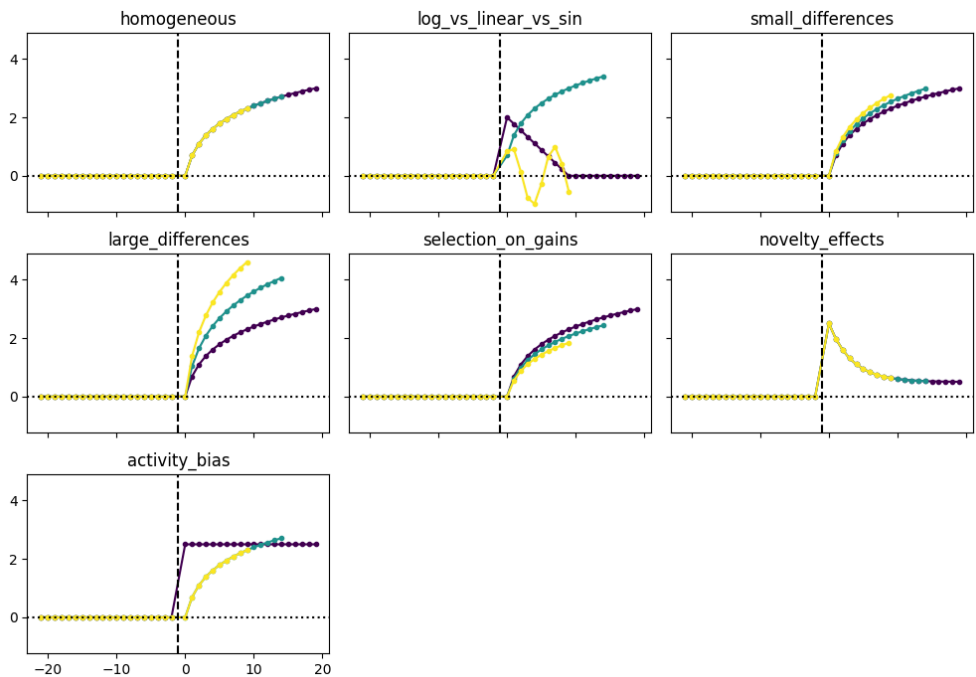

Figure 6: true cohort level effect functions for homogeneous and heterogeneous treatment effects across three cohorts. Earliest-treated cohort is in purple, middle cohort in green, and latest cohort in yellow. 'Homogenous' and 'novelty effects' DGPs have homogeneous treatment effects across cohorts, while all others have heterogeneous treatment effects across cohorts.

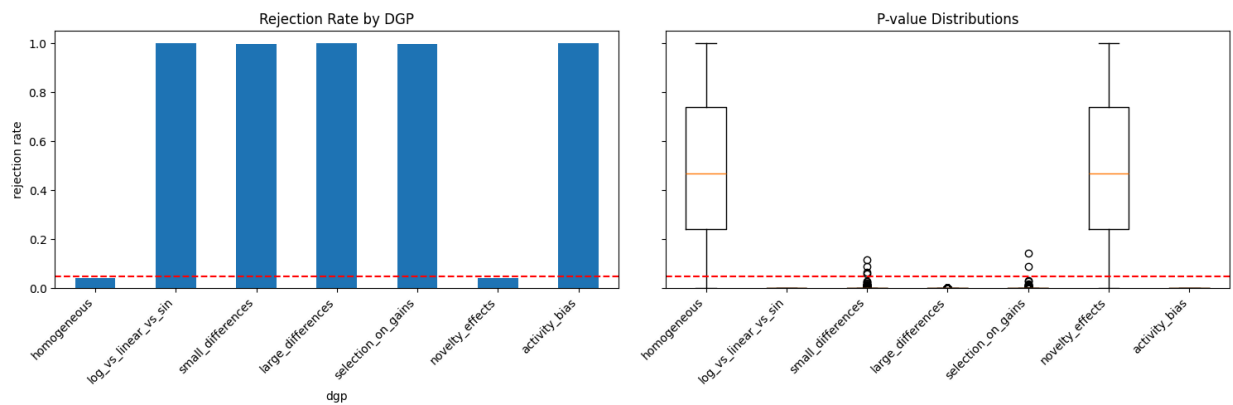

Figure 7: Rejection rates over 1000 replications for the joint test of cohort-level coefficients using an F-test in DGPs from Figure 6

## 5 Conclusion

The two-way fixed effects regression remains a workhorse tool in applied economics despite recent critiques highlighting its potential shortcomings under treatment effect heterogeneity. This paper provides simple diagnostic tests that help researchers determine when TWFE is likely to yield reliable estimates versus when more complex estimators are needed. Our simulation evidence shows these tests have good power to detect problematic patterns of effect heterogeneity while maintaining correct size under the null of homogeneous effects.

The tests we propose are computationally simple and implemented in the pyfixest library and readily implementable in standard statistical software. Since heterogeneity-robust estimators often come with higher variance and computational complexity, the ability to test when they are truly needed helps researchers make principled choices about their estimation strategy. While these tests cannot guarantee TWFE will recover meaningful treatment effects, they provide a practical tool for detecting scenarios where the recent critiques of TWFE are most relevant.